\documentclass[pra,twocolumn,showpacs,preprintnumbers,superscriptaddress]
{revtex4-2}

\usepackage{times}
\usepackage{bm}
\usepackage{float}
\usepackage{graphicx}
\usepackage{amsbsy}
\usepackage{amsmath}
\usepackage{amsfonts}
\usepackage{amsthm}

\begin{document}
	
\theoremstyle{plain}
\newtheorem{theorem}{Theorem}
\newtheorem{lemma}[theorem]{Lemma}
\newtheorem{corollary}[theorem]{Corollary}
\newtheorem{proposition}[theorem]{Proposition}
\newtheorem{conjecture}[theorem]{Conjecture}

\theoremstyle{definition}
\newtheorem{definition}[theorem]{Definition}

\theoremstyle{remark}
	\newtheorem*{remark}{Remark}
	\newtheorem{example}{Example}
	\title{A few entanglement criterion for two-qubit and two-qudit system based on realignment operation}
	\author{Shweta Kalson, Anchal Singh, Satyabrata Adhikari}
	\email{shwetakalson\_2k20mscmat29@dtu.ac.in, \\anchalsingh\_2k20mscmat03@dtu.ac.in, satyabrata@dtu.ac.in} 
	\affiliation{Delhi Technological University, Delhi-110042, Delhi, India}
	
\begin{abstract}
	\begin{center}
		\textbf{Abstract}
	\end{center}
It is known that realignment crierion is necessary but not a sufficient criterion for lower as well as higher dimensional system. In this work, we first consider a two-qubit system and derived the necessary and sufficient condition based on realignment operation for a particular class of two-qubit system. Thus we solved the problem of if and only if condition partially for a particular class of two-qubit state. 
We have shown that the derived necessary and sufficient condition detects two-qubit entangled states, which are not detected by the realignment criterion. Next, we discuss the higher dimensional system and obtained the necessary condition on the minimum singular value of the realigned matrix of $d\otimes d$ dimensional separable states. Moreover, we provide the geometrical interpretation of the derived separability criterion for $d\otimes d$ dimensional system. Furthermore, we show that our criterion may also detect bound entangled state. The entanglement detection criterion studied here is beneficial in the sense that it requires to calculate only minimum singular value of the realigned matrix while on the other hand realignment criterion requires all singular values of the realigned matrix. Thus, our criterion has computational advantage over the realignment criterion.  
\end{abstract}
 \maketitle
	
\section{Introduction}
It is well known that entanglement detection problem \cite{horodecki1,guhne1} is one of the prime problem in quantum information theory. This problem can be considered as NP complete \cite{gurvit} and so all entangled state cannot be detected by just one criterion. Thus, there exist a vast literature in the context of the development of different entanglement detection criterion \cite{horodecki3,horodecki4,nielsen,guhne2,bruss,braunstein,dutta}. Partial transposition (PT) criterion is the first entanglement detection criterion introduced by Peres \cite{peres} and later Horodecki et.al \cite{horodecki2} proved that the PT criterion is necessary and sufficient for $2\otimes 2$ and $2\otimes 3$ dimensional system. Mathematically, If any bipartite state described by the density operator
\begin{center}
    $\rho_{AB}=\begin{pmatrix}
	P & Q \\
	R & S 
\end{pmatrix}$
\end{center}
 where $A,B,C,D$ denote the block matrices then the partial transposition operation performed on $\rho$ may be defined as
\begin{eqnarray}
\rho_{AB}^{T_{B}}=\begin{pmatrix}
	P^{T} & Q^{T} \\
	R^{T} & S^{T} 
\end{pmatrix}
	\label{pt1}
\end{eqnarray}
where $T_{B}$ denote the partial transposition with respect to the subsystem $B$ and $T$ denote the usual transposition operation. In higher dimensional system, PT criterion is only necessary but not sufficient.\\
Like Partial transposition operation, there also exist another operation that permute the elements of the density matrix in a different way and the resulting operation is known as realignment operation \cite{rudolph,kchen}. Realignment operation can be explained as follows: Let us consider a bipartite state 
\begin{center}
 $\rho_{AB}=\sum_{ijkl} \rho_{ijkl}|i j\rangle \langle k l|$   
\end{center}
 in the composite system $H_{A}^{d_{1}} \otimes  H_{B}^{d_{2}}$, where $H_{A}^{d_{1}}$ and  $H_{B}^{d_{2}}$ denoting the Hilbert spaces representing the individual systems $A$ and $B$ of dimension $d_{1}$ and $d_{2}$ respectively. The realignment operation $R$ when acting on the state $\rho_{AB}$ give the output as 
\begin{eqnarray}
	R(\rho_{AB}) = \sum_{ijkl} \rho_{kjil}|k j\rangle \langle i l|
\end{eqnarray}
In matrix notation, the density matrix $\rho_{AB}$ for $2\otimes 2$ system is given by
\begin{eqnarray}
	\rho_{AB}=
	\begin{pmatrix}
		E & F \\
		G & H 
	\end{pmatrix}
\end{eqnarray}
where 
\begin{eqnarray}
	&&E=
	\begin{pmatrix}
		\rho_{1,1} & \rho_{1,2}\\
		\rho^{*}_{1,2} & \rho_{2,2} 	
	\end{pmatrix}, F=
	\begin{pmatrix}
		\rho_{1,3} & \rho_{1,4}\\
		\rho_{2,3} & \rho_{2,4} 	
	\end{pmatrix},\nonumber\\&& G=
	\begin{pmatrix}
		\rho^{*}_{1,3} & \rho^{*}_{2,3}\\
		\rho^{*}_{1,4} & \rho^{*}_{2,4}
	\end{pmatrix}, H=
	\begin{pmatrix}
		\rho_{3,3} & \rho_{3,4}\\
		\rho^{*}_{3,4} & \rho_{4,4}
	\end{pmatrix}
\end{eqnarray}
When realignment operation $R$ performed on $\rho_{AB}$ then the resulting output matrix look like
\begin{eqnarray}
	R(\rho_{AB})=
	\begin{pmatrix}
		R(E)  \\
		R(F) \\
		R(G) \\
		R(H) 
	\end{pmatrix}
\end{eqnarray}
where \begin{eqnarray}
	&&R(E)=
	\begin{pmatrix}
		\rho_{1,1} & \rho_{1,2} & \rho^{*}_{1,2} & \rho_{2,2} 
	\end{pmatrix},\nonumber\\&& R(F)=
	\begin{pmatrix}
		\rho_{1,3} & \rho_{1,4} & \rho_{2,3} & \rho_{2,4} 	
	\end{pmatrix},\nonumber\\&& R(G)=
	\begin{pmatrix}
		\rho^{*}_{1,3} & \rho^{*}_{2,3} & \rho^{*}_{1,4} & \rho^{*}_{2,4}
	\end{pmatrix},\nonumber\\&& R(H)=
	\begin{pmatrix}
		\rho_{3,3} & \rho_{3,4} & \rho^{*}_{3,4} & \rho_{4,4}
	\end{pmatrix}
\end{eqnarray}
Now, we are in a position to review the realignment criterion, which can be stated as follows: If the state $\rho_{AB}$ is separable then $||R(\rho_{AB}) ||_{1} \leq 1$ where $||.||_{1}$ denotes the trace norm and defined by $|| A ||_{1}$ = Tr $\sqrt{A A \dagger}$. Since the right hand side of the inequality is unity in the realignment criterion, which does not depend on the state under investigation so it may be called as state independent realignment criterion.\\
In this work, we focus on developing the alternative form of the Realignment criterion for two reasons: (i) Unlike PT criterion, realignment criterion is necessary but not sufficient even for two-qubit system. This motivates us to investigate for the possible condition based on realignment operation that may be necessary and sufficient for at least two-qubit system. (ii) As the dimension of the composite bipartite system increases, the number of singular values that will be needed to determine the trace norm of the realigned matrix, also increases. by seeing this difficulty, we have developed the separability criterion which is based on only one singular value i.e. minimum singular value of the realigned matrix.\\
The organization of the work is as follows: In section-II, we have derived necessary and sufficient condition based on realignment operation to detect a particular class of two-qubit entangled state. We have provided few examples of two-qubit entangled states which are not detected by realignment criterion but detected by the criterion derived here. In section-III, we have discussed the separability criterion for higher dimensional system and shown that the obtained separability criterion depends only on the minimum singular value of the realigned matrix. The obtained criterion is necessary but not sufficient. Further, we give the geometrical interpretation of the derived separability criterion. Lastly, we conclude in section-IV.  

\section{State dependent realignment criterion}
We should note here an important fact that the realignment criterion is necessary but not sufficient. In this respect, we can ask the following question: Can we derive a necessary and sufficient entanglement condition for a two-qubit state using realignment operation? In this work, we have answered this question partially. We derive here the necessary and sufficient condition for the existence of entanglement in a particular class of a two-qubit system using the realignment operation. The derived criterion is state dependent and hence it can be named as state dependent realignment criterion. The derived separability criterion is important in the sense that it can detect a two-qubit entangled state which is not detected by the realignment criterion.\\
Let us consider a particular class of two-qubit state described by the density operator $\varrho_{AB}$, given by 
\begin{eqnarray}
\varrho_{AB}=
\begin{pmatrix}
\rho_{11} & 0 & 0 & \rho_{14} \\
0 & 0 & 0 & 0 \\
0 & 0 & \rho_{33} & 0 \\
\rho_{14}^{*} & 0 & 0 & \rho_{44} \\
\end{pmatrix},\rho_{11}+\rho_{33}+\rho_{44}=1
\end{eqnarray}
We can always choose the state parameter $\rho_{11}$, $\rho_{14}$, $\rho_{14}^{*}$, $\rho_{33}$ and $\rho_{44}$ in such a way that $\varrho_{AB}$ represent a positive semi-definite matrix. 
The partial transposition of $\varrho_{AB}$ is given by
\begin{eqnarray}
\varrho_{AB}^{T_{B}}=
\begin{pmatrix}
\rho_{11} & 0 & 0 & 0 \\
0 & 0 & \rho_{14} & 0 \\
0 & \rho_{14}^{*} & \rho_{33} & 0 \\
0 & 0 & 0 & \rho_{44} \\
\end{pmatrix}
\end{eqnarray}
where $T_{B}$ denote the partial transposition with respect to the system $B$.\\
The eigenvalues of $\varrho_{AB}^{T_{B}}$ are given as
\begin{eqnarray}
&& \lambda_{1}=\rho_{11}, \lambda_{2}=\rho_{44}\nonumber\\&&
\lambda_{3}=\frac{1}{2}\rho_{33}+\frac{1}{2}\sqrt{\rho_{3,3}^{2}+4|\rho_{14}|^{2}}\nonumber\\&& \lambda_{4}=\frac{1}{2}\rho_{33}-\frac{1}{2}\sqrt{\rho_{3,3}^{2}+4|\rho_{14}|^{2}}
\label{partrans1}
\end{eqnarray}
It can be observed that the eigenvalues $\lambda_{1}$, $\lambda_{2}$ and $\lambda_{3}$ will always be positive. The eigenvalue $\lambda_{4}$ will only be positive when $|\rho_{14}|=0$. Thus, $\varrho_{AB}^{T_{B}}$ has negative eigenvalues only if $|\rho_{14}|\neq 0$ and hence for any non-zero state parameter $|\rho_{14}|$, the state $\varrho_{AB}$ represent an entangled state.\\
After performing realignment operation on the state $\rho_{AB}$, the state reduces to 
\begin{eqnarray}
R(\varrho_{AB})=
\begin{pmatrix}
\rho_{11} & 0 & 0 & 0 \\
0 & \rho_{14} & 0 & 0 \\
0 & 0 & \rho_{14}^{*} & 0 \\
\rho_{33} & 0 & 0 & \rho_{44} \\
\end{pmatrix}
\end{eqnarray}
If $s_{1}$, $s_{2}$, $s_{3}$, $s_{4}$ denote the singular values of $R(\varrho_{AB})$ then the singular values are given by 
\begin{eqnarray}
&& s_{1}=\sqrt{\frac{1}{2}[f(\rho_{11},\rho_{33},\rho_{44})+\sqrt{f(\rho_{11},\rho_{33},\rho_{44})^2-4(\rho_{11}\rho_{44})^2}]},\nonumber\\&&
s_{2}=\sqrt{\frac{1}{2}[f(\rho_{11},\rho_{33},\rho_{44})-\sqrt{f(\rho_{11},\rho_{33},\rho_{44})^2-4(\rho_{11}\rho_{44})^2}]},\nonumber\\&& 
s_{3}=\frac{|\rho_{1,4}|}{2}, s_{4}=\frac{|\rho_{1,4}|}{2} 
\label{singval}
\end{eqnarray}
where $f(\rho_{11},\rho_{33},\rho_{44})=\rho_{11}^{2}+\rho_{33}^{2}+\rho_{44}^{2}$.\\
The trace norm of $R(\varrho_{AB})$ can be calculated as
\begin{eqnarray}
||R(\varrho_{AB}) ||_{1}&=& s_{1}+s_{2}+s_{3}+s_{4}\nonumber\\&=&
s_{1}+s_{2}+|\rho_{14}|
\label{tracenorm1}
\end{eqnarray}
Applying $AM-GM$ inequality on $s_{1}$ and $s_{2}$, we get
\begin{eqnarray}
s_{1}+s_{2}\geq 2(s_{1}s_{2})^{\frac{1}{2}}=2\sqrt{\rho_{11}\rho_{44}}
\label{am-gm1}
\end{eqnarray}
Using (\ref{am-gm1}) in (\ref{tracenorm1}), we get
\begin{eqnarray}
||R(\varrho_{AB}) ||_{1}&\geq& 2\sqrt{\rho_{11}\rho_{44}}+|\rho_{14}|
\nonumber\\&=& 2\sqrt{\rho_{11}\rho_{44}}+\sqrt{-\lambda_{3}\lambda_{4}}
\label{tracenorm2}
\end{eqnarray}
Here, $\lambda_{4}$ denote the minimum eigenvalue of partial transposed matrix of $\varrho_{AB}$ and from partial transposition criterion, we can say that $\lambda_{4}$ will be negative if and only if the state $\varrho_{AB}$ is an entangled state. Since it has also been shown in \cite{rana} that all eigenvalues of partial transposed matrix lying in $[-\frac{1}{2},1]$  so, $\lambda_{4}\in [-\frac{1}{2},0]$. Thus, we have the following inequality
\begin{eqnarray}
\lambda_{4} < 0 \implies \frac{\rho_{33}}{2} < \frac{1}{2}\sqrt{\rho_{33}^{2}+4|\rho_{14}|^{2}}
\label{mineig1}
\end{eqnarray}
Using (\ref{mineig1}) in (\ref{tracenorm2}), we get
\begin{eqnarray}
||R(\varrho_{AB}) ||_{1}&\geq&  2\sqrt{\rho_{11}\rho_{44}}+\nonumber\\&&\sqrt{-\rho_{33}(\frac{\rho_{33}}{2}-\frac{1}{2}\sqrt{\rho_{33}^{2}+4|\rho_{14}|^{2}})}
\label{tracenorm3}
\end{eqnarray}
Now, we are in a position to state the following theorem for the particular class of two-qubit state as:\\ \\
\textbf{Theorem-1:} Let us consider a class of two-qubit state described by the density operator $\varrho_{AB}=\rho_{11}|00 \rangle\langle 00|+\rho_{14}|00 \rangle\langle 11|+\rho_{33}|10\rangle\langle 10|+\rho_{14}^{*}|11\rangle\langle 00|+\rho_{44}|11\rangle\langle 11|]$. The state $\varrho_{AB}$ is entangled if and only if 
\begin{eqnarray}
||R(\varrho_{AB}) ||_{1}&\geq&  2\sqrt{\rho_{11}\rho_{44}}+\nonumber\\&&\sqrt{-\rho_{33}(\frac{\rho_{33}}{2}-\frac{1}{2}\sqrt{\rho_{33}^{2}+4|\rho_{14}|^{2}})}
\label{th1}
\end{eqnarray}
\subsection{A few states which are not detected by realignment criterion but detected by our criterion}
In this section, we will discuss about few states which are not detected by realignment criterion but they may be detected either by using Theorem-1 or Corollary-1.\\ \\
\textbf{Example-1:} Let us consider a state $\rho_{1}$, which is given by \cite{rudolph}
\begin{eqnarray}
\rho_{1}=
\begin{pmatrix}
\frac{5}{8} & 0 & 0 & \frac{1}{32} \\
0 & 0 & 0 & 0 \\
0 & 0 & \frac{1}{8} & 0 \\
\frac{1}{32} & 0 & 0 & \frac{1}{4} \\
\end{pmatrix}
\end{eqnarray}
The partial transposition of the density matrix $\rho_{1}$ is given by
\begin{eqnarray}
\rho_{1}^{T_{B}}= 
\begin{pmatrix}
\frac{5}{8} & 0 & 0 & 0 \\
0 & 0 & \frac{1}{32} & 0 \\
0 & \frac{1}{32} & \frac{1}{8} & 0 \\
0 & 0 & 0 & \frac{1}{4}\\
\end{pmatrix}
\end{eqnarray}
The eigenvalues of partial transposed matrix $\rho_{1}^{T_{B}}$ are given by
\begin{eqnarray}
&& \lambda_{1}^{(1)}= \frac{5}{8} , \lambda_{2}^{(1)}= \frac{1}{4} ,\lambda_{3}^{(1)}= \frac{1}{32}(2+\sqrt{5})\nonumber\\&&
,\lambda_{4}^{(1)}= \frac{1}{32}(2-\sqrt{5})
\label{ex11}
\end{eqnarray}\\
Since one eigenvalue of $\rho_{1}^{T_{B}}$ is negative so, the state $\rho_{1}$ is an entangled state.
The realigned matrix of $\rho_{1}$ is given by
\begin{eqnarray}
R(\rho_{1})=
\begin{pmatrix}
\frac{5}{8} & 0 & 0 & 0 \\
0 & \frac{1}{32} & 0 & 0 \\
0 & 0 & \frac{1}{32} & 0 \\
\frac{1}{8} & 0 & 0 & \frac{1}{4} \\
	\end{pmatrix}
\end{eqnarray}
The trace norm of $R(\rho_{1})$ are given by
\begin{eqnarray}
||R(\rho_{1}) ||_{1}= 0.9464
\label{tracenorm13}
\end{eqnarray}
We can observe that $||R(\rho_{1}) ||_{1}<1$ and thus the realignment criterion failed to conclude whether the state $\rho_{1}$ is entangled or not. Now, our task is to test for the entanglement of the state $\rho_{1}$ using theorem-1.\\ 
The LHS and RHS of the inequality in theorem-1 gives
\begin{eqnarray}
&&LHS=||R(\rho_{1}) ||_{1}= 0.9464 \nonumber\\&&
RHS= 2\sqrt{\rho_{11}\rho_{44}}\nonumber\\&&
+\sqrt{-\rho_{33}(\frac{\rho_{33}}{2}-\frac{1}{2}\sqrt{\rho_{33}^{2}+4|\rho_{14}|^{2}})}= 0.8207
\label{rhs1}
\end{eqnarray}
Therefore, Theorem-1 is verified for the state $\rho_{1}$ and thus it can be concluded that $\rho_{1}$ is an entangled state.\\ \\
\textbf{Example-2:} Let us consider another state described by the state $\rho_{2}$
\begin{eqnarray}
\rho_{2}= \frac{1}{2}
\begin{pmatrix}
\frac{7}{6} & 0 & 0 & \frac{1}{14} \\
0 & 0 & 0 & 0 \\
0 & 0 & \frac{1}{3} & 0 \\
\frac{1}{14} & 0 & 0 & \frac{1}{2} \\
\end{pmatrix}
\end{eqnarray}
The partial transposition of the density matrix $\rho_{2}$ is given by
\begin{eqnarray}
\rho_{2}^{T_{B}}= \frac{1}{2}
\begin{pmatrix}
\frac{7}{6} & 0 & 0 & 0 \\
0 & 0 & \frac{1}{14} & 0 \\
0 & \frac{1}{14} & \frac{1}{3} & 0 \\
0 & 0 & 0 & \frac{1}{2} \\
\end{pmatrix}
\end{eqnarray}
The eigenvalues of $\rho_{2}^{T_{B}}$ are given by $\lambda_{1}^{(2)}=0.2500$, $\lambda_{2}^{(2)}=0.5833$, $\lambda_{3}^{(2)}=0.1740$, $\lambda_{4}^{(2)}=-0.0073$. By partial transposition criterion, the state $\rho_{2}^{T_{B}}$ is an entangled state.\\
The realigned matrix $R(\rho_{2})$ is given by
\begin{eqnarray}
R(\rho_{2})=\frac{1}{2}
\begin{pmatrix}
\frac{7}{6} & 0 & 0 & 0 \\
0 & \frac{1}{14} & 0 & 0 \\
0 & 0 & \frac{1}{14} & 0 \\
\frac{1}{3} & 0 & 0 & \frac{1}{2}\\
\end{pmatrix}
\end{eqnarray}
The trace norm of the realigned matrix $R(\rho_{2})$ is found out to be $||R(\rho_{2})||_{1}=0.9213$. Therefore, in this case also realignment criterion does not detect the entangled state $\rho_{2}$.\\
To detect the entangled state $\rho_{2}$, let us now use our criterion given in (\ref{th1}). The RHS of (\ref{th1}) is given by the quantity  $2\sqrt{\rho_{11}\rho_{44}}+\sqrt{-\rho_{33}(\frac{\rho_{33}}{2}-\frac{1}{2}\sqrt{\rho_{33}^{2}+4|\rho_{14}|^{2}})}$ and it can be calculated for the state $\rho_{2}$ as $0.795$. Thus, our criterion (\ref{th1}) is satisfied and hence the state $\rho_{2}$ is verified to be an entangled state. 
\section{Separability criterion for $d\otimes d$ dimensional system based on realignment operation}
In this section, we will derive few separability criterion for $d\otimes d$ dimensional system using two other existing separability criterion, namely, (i) realignment criterion \cite{rudolph} and (ii) structural physical approximation of partial transposition (SPA-PT) criterion \cite{horodeckiekert}. The derived criterion provides only necessary condition for a state under probe to be a separable state.\\

\subsection{Separability criterion in terms of minimum singular value of realigned matrix}
Let us consider an arbitrary $d \otimes d$ dimensional bipartite state shared by two distant partners Alice (A) and Bob(B), which is described by the density operator $\rho_{AB}$. We can now present the separability criterion given in the following theorem:\\ \\
\textbf{Theorem-2:} If $R(\rho_{AB})$ denote the realigned form of the density matrix $\rho_{AB}$ and if $\rho_{AB}$ is separable then
\begin{eqnarray}
s_{min}[R(\rho_{AB})]< \frac{1}{d^{2}}
\label{th2}
\end{eqnarray}
where $s_{min}[R(\rho_{AB})]$ denotes the minimum singular value of $R(\rho_{AB})$.\\
Proof: If $\tilde{\rho}_{AB}$ represents the SPA-PT of $\rho_{AB}^{T_{B}}$ 
then all eigenvalues of $\tilde{\rho}_{AB}$ are positive. Thus $\lambda_{min}[\tilde{\rho}_{AB}]$ is also positive. We can now apply $AM-GM$ inequality on the two positive quantities $||R(\rho_{AB})||_{1}$ and $\lambda_{min}[\tilde{\rho}_{AB}]$ and we get
\begin{eqnarray}
\frac{||R(\rho_{AB})||_{1}+\lambda_{min}[]\tilde{\rho}_{AB}]}{2}&\geq& [||R(\rho_{AB})||_{1}\lambda_{min}[\tilde{\rho}_{AB}]]^{\frac{1}{2}}\nonumber\\&=& [\sum_{i=1}^{d^{2}}s_{i}[R(\rho_{AB})]\lambda_{min}[\tilde{\rho}_{AB}]]^{\frac{1}{2}}\nonumber\\&\geq& d[s_{min}[R(\rho_{AB})]\lambda_{min}[\tilde{\rho}_{AB}]]^{\frac{1}{2}}\nonumber\\
\label{pf21}
\end{eqnarray}
Since, it is given that the state $\rho_{AB}$ is separable so, from the realignment criterion we have
\begin{eqnarray}
||R(\rho_{AB})||_{1}\leq 1
\label{pf22}
\end{eqnarray}
Using (\ref{pf22}) and after a little bit of simplification, the inequality (\ref{pf21}) reduces to
\begin{eqnarray}
[\lambda_{min}[\tilde{\rho}_{AB}]]^{2}+2(1-2d^{2}s_{min}[R(\rho_{AB})])\lambda_{min}[\tilde{\rho}_{AB}]+1\geq 0 \nonumber\\
\label{pf23}
\end{eqnarray}
The expression given in (\ref{pf23}) represents a quadratic equation in $\lambda_{min}[\tilde{\rho}_{AB}]$ and since it is non-negative so the discriminant of the quadratic expression must be negative. Thus, we have
\begin{eqnarray}
s_{min}[R(\rho_{AB})]<\frac{1}{d^{2}}
\label{pf24}
\end{eqnarray}
Hence proved.\\ \\
\textbf{Corollary-1}: If $\rho_{AB}\in H_{A}^{d}\otimes H_{B}^{d}$, where $H_{A}^{d}$ and  $H_{B}^{d}$ denoting the Hilbert spaces representing the individual systems $A$ and $B$ of dimension $d$ each respectively and if the inequality (\ref{th2}) is violated for any state $\rho_{AB}$ then the state $\rho_{AB}$ is an entangled state.\\
\subsection{Geometrical Interpretation of separability criterion}
In this section, we study the geometrical interpretation of the separability condition. To start with, we assume that the state $\rho_{AB}$ is separable. We can then recall the separability condition (\ref{pf23}) in terms of the non-negativity of the quadratic expression, which can be re-expressed as
\begin{eqnarray}
(x+\frac{A}{2})^{2}=y+\frac{A^{2}-4}{4}
\label{g1}
\end{eqnarray}
where $x=\lambda_{min}[\tilde{\rho}_{AB}]$, $A=2(1-2d^{2}s_{min}[R(\rho_{AB})])$ and $y=[\lambda_{min}[\tilde{\rho}_{AB}]]^{2}+2(1-2d^{2}s_{min}[R(\rho_{AB})])\lambda_{min}[\tilde{\rho}_{AB}]+1$.\\
The equation (\ref{g1}) represents a parabola with vertex at $(-\frac{A}{2},\frac{4-A^{2}}{4})$. 
\begin{figure}[h!]
    \centering
    \includegraphics[width=10cm]{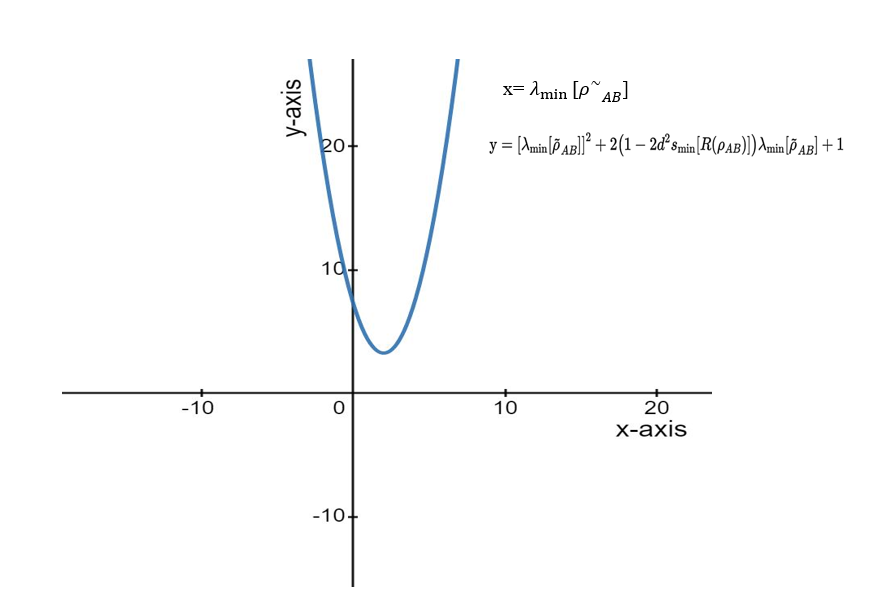} 
    \caption{Geometrically, the states lying inside the parabola are entangled states while all separable states are lying outside the parabola. But there exist entangled states that are lying even outside the parabola.}
    \label{fig:my_label}
\end{figure}
Since, the state $\rho_{AB}$ is separable so (\ref{th2}) holds and thus the vertex of the parabola lie at the first quadrant of $x-y$ plane. Hence, the separability criterion given in theorem-1 can be interpreted as "all separable states are lying outside the parabola". Since the derived separability criterion is only necessary but not sufficient so there exist entangled states that may lie outside the parabola. Further, we may note that the states lying inside the parabola are entangled states and this interpretation may be claimed from corollary-1.
\subsection{Examples} 
In this section, we will take few examples to verify the result given in Theorem-2 and Corollary-1. To accomplish this task, we first consider negative partial transpose entangled state and then we consider positive partial transpose entangled state or bound entangled state.
\subsubsection{Detection of negative partial transpose entangled state (NPTES)}
\textbf{Example-3} Let us consider a $3 \otimes 3$ dimensional state described by the density operator $\rho_{3}$
\begin{eqnarray}
\rho_{3}= 
\begin{pmatrix}
P_{1} & Q_{1} & R_{1} \\
Q_{1}^{\dagger} & Q_{2} & R_{2} \\
R_{1}^{\dagger} & R_{2}^{\dagger} & R_{3} \\
\end{pmatrix}, 0\leq f \leq 1
\label{rho3}
\end{eqnarray}
where,\\
$P_{1}= 
\begin{pmatrix}
\frac{2+6f}{24} & 0 & 0 \\
0 & \frac{1-f}{8} & 0 \\
0 & 0 & \frac{1-f}{8} \\
\end{pmatrix}$; $Q_{1}= 
\begin{pmatrix}
0 & \frac{9f-1}{24} & 0 \\
0 & 0 & 0 \\
0 & 0 & 0 \\
\end{pmatrix}$;\\$R_{1}= 
\begin{pmatrix}
0 & 0 & \frac{9f-1}{24} \\
0 & 0 & 0 \\
0 & 0 & 0 \\
\end{pmatrix}$; $Q_{2}= 
\begin{pmatrix}
\frac{1-f}{8} & 0 & 0 \\
0 & \frac{2+6f}{24} & 0 \\
0 & 0 & \frac{1-f}{8} \\
\end{pmatrix}$;\\ $R_{2}= 
\begin{pmatrix}
0 & 0 & 0 \\
0 & 0 & \frac{9f-1}{24} \\
0 & 0 & 0 \\
\end{pmatrix}$; $R_{3}= 
\begin{pmatrix}
\frac{1-f}{8} & 0 & 0 \\
0 & \frac{1-f}{8} & 0 \\
0 & 0 & \frac{2+6f}{24} \\
\end{pmatrix}$\\
The state $\rho_{3}$ is separable for $0\leq f \leq \frac{1}{3}$ and entangled for $\frac{1}{3}< f \leq 1$.
The realigned form of the density matrix $\rho_{3}$ is denoted by $R(\rho_{3})$ and it is given by
\begin{eqnarray}
R(\rho_{3})= 
	\begin{pmatrix}
		P_{1}^{(R)} & Q_{1}^{(R)} & R_{1}^{(R)} \\
		(Q_{1}^{\dagger})^{(R)} & Q_{2}^{(R)} & R_{2}^{(R)} \\
		(R_{1}^{\dagger})^{(R)} & (R_{2}^{\dagger})^{(R)} & R_{3}^{(R)} \\
	\end{pmatrix}, 0\leq f \leq 1
	\label{rho3}
\end{eqnarray}
where,\\ $P_{1}^{(R)}= 
\begin{pmatrix}
	\frac{2+6f}{24} & 0 & 0 \\
	0 & \frac{9f-1}{24} & 0 \\
	0 & 0 & \frac{9f-1}{24} \\
\end{pmatrix}$; $Q_{1}^{(R)}= 
\begin{pmatrix}
	0 & \frac{1-f}{8} & 0 \\
	0 & 0 & 0 \\
	0 & 0 & 0 \\
\end{pmatrix}$; $R_{1}^{(R)}= 
\begin{pmatrix}
	0 & 0 & \frac{1-f}{8} \\
	0 & 0 & 0 \\
	0 & 0 & 0 \\
\end{pmatrix}$; $Q_{2}^{(R)}= 
\begin{pmatrix}
	\frac{9f-1}{24} & 0 & 0 \\
	0 & \frac{2+6f}{24} & 0 \\
	0 & 0 & \frac{9f-1}{24} \\
\end{pmatrix}$; $R_{2}^{(R)}= 
\begin{pmatrix}
	0 & 0 & 0 \\
	0 & 0 & \frac{1-f}{8} \\
	0 & 0 & 0 \\
\end{pmatrix}$; $R_{3}^{(R)}= 
\begin{pmatrix}
	\frac{9f-1}{24} & 0 & 0 \\
	0 & \frac{9f-1}{24} & 0 \\
	0 & 0 & \frac{2+6f}{24} \\
\end{pmatrix}$\\
The minimum singular value of $R(\rho_{3})$ is given by
\begin{eqnarray}
s_{min}(R(\rho_{3}))= \frac{9f-1}{24} 
\label{singvalRrho3}
\end{eqnarray}
Therefore, applying Theorem-1, we find that the state $\rho_{3}$ is separable when the state parameter $f \in [0,\frac{1}{3}]$. Further, using corollary-1, we conclude that the state $\rho_{3}$ is entangled if $f>\frac{11}{27}\approx 0.4074$.\\
\begin{figure}[h!]
    \centering
    \includegraphics[width=10cm]{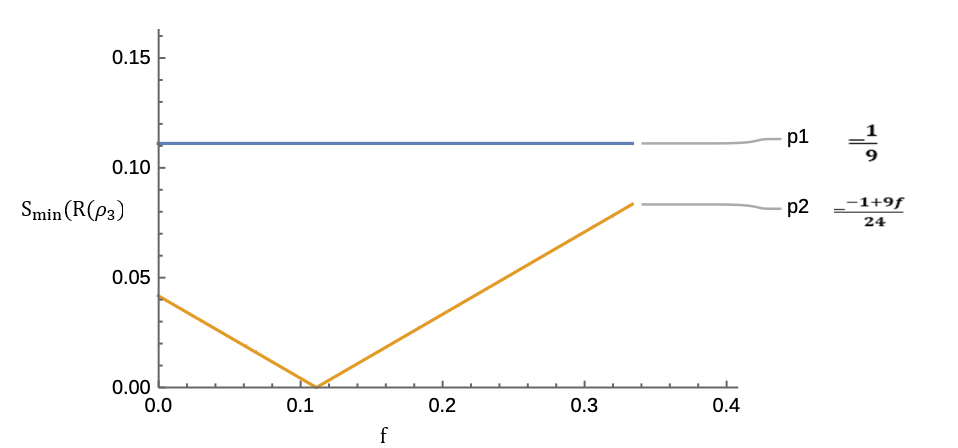} 
    \includegraphics[width=10cm]{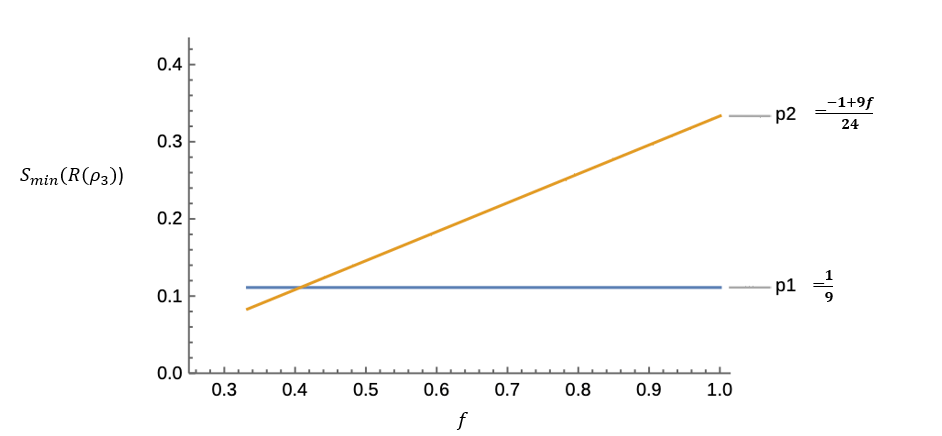}
    \caption{(a) In this figure, $s_{min}(R(\rho_{3}))\leq \frac{1}{9}$ when $0\leq f \leq \frac{1}{3}$ and thus $\rho_{3}$ represent a separable state (b) In this figure, $s_{min}(R(\rho_{3})) >  \frac{1}{9}$ when $ \frac{11}{27} < f \leq 1$ and thus $\rho_{3}$ represent an entangled state $\frac{1}{3} < f \leq 1$}. 
    \label{fig:my_label}
\end{figure}

\textbf{Example-4} Let us consider another free entangled state (NPTES) which is described by the density operator $\rho_{\beta}$ \cite{phorodecki}
\begin{eqnarray}
\rho_{\beta}=\frac{2}{7}|\psi^{+}\rangle\langle \psi^{+}|+\frac{\beta}{7}\sigma^{+}+\frac{5-\beta}{7}\sigma^{-}, 4\leq \beta \leq 5
\label{rhobeta}
\end{eqnarray}
where $|\psi^{+}\rangle=\frac{1}{\sqrt{3}}(|00\rangle+|11\rangle+|22\rangle)$,
$\sigma^{+}=\frac{1}{3}|01\rangle\langle 01|+|12\rangle\langle 12|+ |20\rangle\langle 20|$, $\sigma^{-}=\frac{1}{3}|10\rangle\langle 10|+|21\rangle\langle 21|+ |02\rangle\langle 02|$.\\
The state $\rho_{\beta}$ is NPTES for $4\leq \beta \leq 5$ \cite{phorodecki}.\\
The minimum singular value of $\rho_{\beta}$ is given by
\begin{eqnarray}
s_{min}(\rho_{\beta})= \frac{1}{21}\sqrt{3\beta^{2}-15\beta+19}
\label{singvalrhobeta}
\end{eqnarray}
It can be easily found that $s_{min}(\rho_{\beta})$ is always greater than $\frac{1}{9}$ for $4\leq \beta \leq 5$. Therefore, the state $\rho_{\beta}$ violate Theorem-1. Thus using corollary-1, we can say that the state $\rho_{\beta}$ is entangled, which is indeed true.\\
\begin{figure}[h!]
    \centering
    \includegraphics[width=10cm]{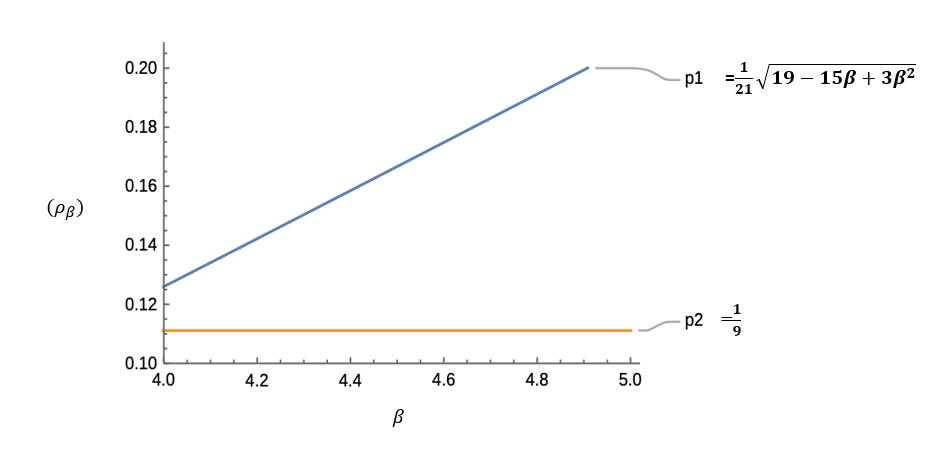} 
    \caption{Figure shows that the state $\rho_{\beta}$ represent an entangled state when $4\leq \beta \leq 5$}
    \label{fig:my_label}
\end{figure}

\subsubsection{Detection of positive partial transpose entangled states (PPTES)} 
\textbf{Example-5} Let us again recall the state $\rho_{\beta}$ given by (\ref{rhobeta}). Considering the different range of the state parameter $\beta$, the state $\rho_{\beta}$ can be re-expressed as
\begin{eqnarray}
\rho_{\beta}^{(1)}=\frac{2}{7}|\psi^{+}\rangle\langle \psi^{+}|+\frac{\beta}{7}\sigma^{+}+\frac{5-\beta}{7}\sigma^{-}, 2\leq \beta \leq 4
\label{rhobeta1}
\end{eqnarray}
\begin{figure}[h!]
	\centering
	\includegraphics[width=9cm]{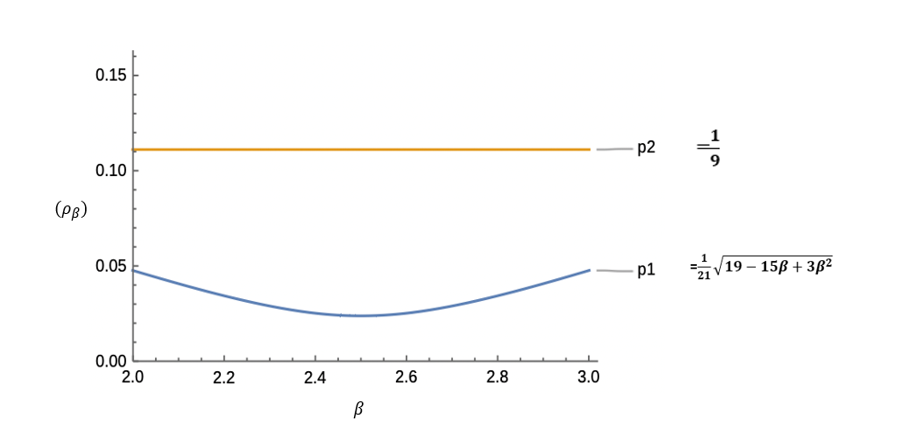} 
	\includegraphics[width=9cm]{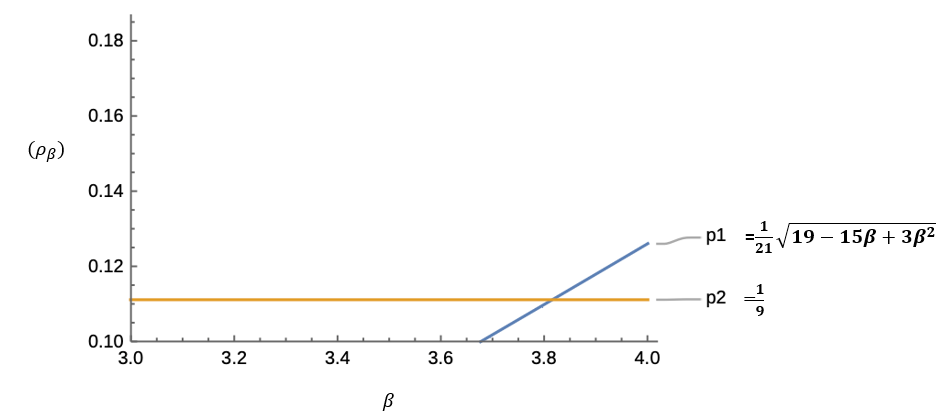}
	\caption{(a) Figure shows that the state $\rho_{\beta}$ represent a separable state when  $2 \leq \beta \leq 3$ (b) Figure shows that the state $\rho_{\beta}$ represent a bound entangled state when $3.815 \leq \beta \leq 4$}
	\label{fig:my_label}
\end{figure}
In \cite{phorodecki}, it has been shown that the state $\rho_{\beta}^{(1)}$ is separable for $2\leq \beta \leq 3$ and PPTES for $3 < \beta \leq 4$. We can verify this result using our criterion which is based on minimum singular value of the realigned matrix of $\rho_{\beta}^{(1)}$. The minimum singular value of the realigned matrix $R(\rho_{\beta}^{(1)})$ is denoted by $s_{min}(R(\rho_{\beta}^{(1)}))$ and it is given by (\ref{singvalrhobeta}).\\
The graph is plotted for minimum singular value $s_{min}(R(\rho_{\beta}^{(1)}))$ and we find that $s_{min}(R(\rho_{\beta}^{(1)}))$ satisfies Theorem-1 for $2\leq \beta \leq 3$. Thus, the state $\rho_{\beta}^{(1)}$ is again verified to be separable using our criterion given in Theorem-1. Moreover, corollary-1 detects PPTES when $\beta \in [3.815,4]$.
\section{Conclusion}
To summarize, we have reviewed the realignment criterion which is  necessary but not sufficient condition for both lower as well as higher dimensional system. To fill this gap for lower dimensional system, we have considered a particular class of two-qubit state and derived a state dependent criterion based on the realignment operation. Our criterion is state dependent in the sense that the RHS of the inequality derived here as an entanglement criterion depends on the state. We may call it as state dependent realignment criterion. We should note here that the RHS of the inequality appeared in the realignment criterion is unity and thus it does not depend on the state. We have shown that the state dependent realignment criterion is necessary and sufficient for the considered particular class of two-qubit state. It is now open for the possible extension of the state dependent realignment criterion for any arbitrary bipartite two qubit state and arbitrary bipartite higher dimensional system. Moreover, to reduce the computational complexity in the calculation of the trace norm of the realigned matrix of the higher dimensional system which is required in realignment criterion, we provide here another separability condition which is based on the minimum singular value of the realignment of the $d\otimes d$ dimensional density matrix. This separability condition is necessary but not sufficient. We also study the geometrical interpretation of the derived separability condition. We have also shown that our criterion has the capability of detecting the bound entangled state.

\section{Data Availability Statement}
Data sharing not applicable to this article as no datasets were generated or analysed during the current study.

\end{document}